\documentstyle[floats,epsf,prd,aps]{revtex}

\newcommand{\be}{\begin{equation}}
\newcommand{\ee}{\end{equation}}

\title{Casimir energy of massive MIT fermions in a Bohm-Aharonov
background}

\author{C.~G.~Beneventano\thanks{Fellow FOMEC-UNLP (Argentina)} \and}

\address{Departamento de F\'{\i}sica, Facultad de Ciencias Exactas,
Universidad Nacional de La Plata \\ C.C. 67 (1900) La Plata, Argentina}

\author{M.~De~Francia\thanks{External Fellow of CONICET (Argentina)} \and
K.~Kirsten \and}

\address{Department of Physics and Astronomy, The University of Manchester
\\ Theory Group, Schuster Laboratory, Manchester M13 9PL, England}

\author{E.~M.~Santangelo\thanks{Member of CONICET (Argentina)}}

\address{Departamento de F\'{\i}sica, Facultad de Ciencias Exactas,
Universidad Nacional de La Plata \\ C.C. 67 (1900) La Plata, Argentina}

\date{October 19, 1999}

\begin{document}
\draft

\maketitle

\begin{abstract}

We study the effect of a background flux string on the vacuum energy of
massive Dirac fermions in 2+1 dimensions confined to a finite spatial
region through MIT boundary conditions. We treat two admissible
self-adjoint extensions of the Hamiltonian and compare the results. In
particular, for one of these extensions, the Casimir energy turns out to
be discontinuous at integer values of the flux.

\end{abstract}


\section{Introduction}
\label{sec-int}

In a by now classic paper, Y. Aharonov and D. Bohm \cite{Bohm} pointed out
that a locally trivial vector potential can, however, give rise to
observable effects in a nontrivial topology. Since then, the relevance of
Aharonov-Bohm scenarios both in particle physics and condensed matter was
recognized.

More recently, much attention has been paid to the inclusion of spin,
mainly in connection with the interaction of cosmic strings with matter
\cite{AlRuWil1989,Russel1988,AlWil1989,Hagen,Flekkoy}. In this context,
the need to consider self-adjoint extensions of the radial Dirac
Hamiltonian was understood \cite{GerbertJackiw,Gerbert}.

Thereafter, the vacuum properties of Dirac fields in the background of a
singular
\cite{Niemi1984,Sitenko1996gd,sitenko97,sitenko96,sitenkohep97,Moroz} as
well as extended \cite{fry,Bordag:1998} magnetic flux string have been
extensively studied. Scalar field and the electromagnetic field have been
considered as well; some pertinent references are
\cite{bretvik,parker,Bordag:1991,scandurra}.

As is well known, the presence of background fields modifies the energy
spectrum, thus giving rise to a nontrivial vacuum, or Casimir, energy
\cite{casimir,Plunien}. Furthermore, the Casimir energy is altered by the
presence of boundaries, and the consequent imposition of boundary
conditions on the quantum fields. For Dirac fields, many examples of both
situations have been studied in the literature (see, for
instance,\cite{Bordag:1998,nos1,nos2,Elizalde:1997we}).

In particular, the combined effect of a classical magnetic fluxon and MIT
boundary conditions on the vacuum energy of a massless Dirac field in 2+1
dimensions was treated in \cite{Leseduarte1996}. There, one of the
possible self-adjoint extensions of the radial Hamiltonian was considered.

In this paper, we will consider the more realistic case of massive
fermions. In this context, it is important to mention that, despite
general belief, in the presence of curved boundaries the effect of a mass
is not exponentially small \cite{Bordag:1996ma} as it is for parallel
plates \cite{Plunien}. On the contrary, in some situations it might even
lead to a sign change in the Casimir force and is by no means negligible
\cite{Bordag:1996ma}. In the present context, we will see that properties
like existence of a minimum or continuity of the Casimir energy as a
function of the flux depend on the mass which underlines its crucial
importance. Moreover, we will analyze two possible self-adjoint
extensions, both known to be compatible with the presence of a Dirac delta
magnetic field at the origin \cite{manuel}. The resulting Casimir energies
are quite different showing that the one-parameter family of self-adjoint
extensions describes nontrivial physics in the core, see also
\cite{Gerbert}.

The organization of the paper is as follows.

In Sec.~\ref{sec-setting} we summarize the generalities of the model.

In Sec.~\ref{sec-origen}, we present a discussion of self-adjoint
extensions distinguished by the behaviour of the wave function at the
origin. We determine the energy eigenfunctions corresponding to two
different cases of these extensions. The first one is a minimal divergence
extension. As shown in \cite{nos1,nos2}, it arises when imposing
Atiyah-Patodi-Singer boundary conditions
\cite{aps,aps2,aps3,spectral,Ma,Ninomiya} at a finite radius, then
shrinked to zero (an idea first suggested in \cite{Poly}). The second one
follows from the zero radius limit of a cylindrical flux shell
\cite{AlRuWil1989,Hagen,Flekkoy}.

In Sec.~\ref{sec-bounded}, the implicit equations for the energy spectrum
are found in both cases, once the theory is confined to a circle of radius
R and MIT conditions (see, for instance, \cite{mit2} and references
therein) are imposed at the exterior boundary. The expression of the
Casimir energy for both types of behaviour at the origin is given in the
framework of the $\zeta$-function regularization
\cite{zeta,Dowker:1976tf,eliz-book1,eliz-book2}.

In Sec.~\ref{sec-eval}, we evaluate the vacuum energies, following the
methods developed in
\cite{Leseduarte1996,Romeo1993,Romeo1994,Bordag:1996gm,Bordag:1996ma}.

Finally, Sec.~\ref{sec-discusion} contains a discussion of the results.

\section{Setting of the problem}
\label{sec-setting}

We study the Dirac equation for a massive particle in $2+1$ dimensional
Minkowski space
\begin{equation}
\left( i \not\!\partial - \not\!\! A - m \right)\Psi = 0
\label{ec-1}
\end{equation}
in the presence of a flux string located at the origin, i.e.,
\begin{equation}
\vec{H} = \vec{\nabla} \wedge \vec{A} = \frac{\kappa}{r} \delta
(r) \check{e}_z\,,
\end{equation}
where $\kappa=\frac{\Phi}{2\pi}$ is the reduced flux.

We assume the flux string to be radially symmetric; so, a gauge can be
chosen such that the (covariant) vector potential is given by
\begin{equation}
A_\theta\left(r\right) =-\frac{\kappa}{r}\,,\qquad\text{for}\quad r>0\,.
\end{equation}

We will consider the chiral representation for the Dirac matrices
\be
\gamma^0 = \rho_3 \otimes \sigma_3, \qquad \gamma^1 = i \rho_3
\otimes \sigma_2, \qquad \gamma^2= -i \rho_3 \otimes \sigma_1,
\label{dirac} \ee which, together with
\be
\gamma^3 = i \rho_2 \otimes \sigma_0 \,,\ee give a closed Clifford
algebra.

Then, the  eigenvalue equation for the Dirac Hamiltonian takes the
form
\be
\left(
\begin{array}{cc}
H_+ & 0 \\ 0 & H_-
\end{array} \right) \Psi_E = E \, \Psi_E\,,
\label{hache}\ee
where the two-by-two blocks are given by
\be
H_\pm = \left(
\begin{array}{cc}
\mp m & L^\dag \\ L & \pm m
\end{array} \right),
\ee
and we have introduced $L = - i e^{ i \theta} \left( -
\partial_r + B \right)$, $L^\dag = i e^{- i \theta} \left(
\partial_r + B \right)$, $B=-\frac{i}{r} \partial_\theta
-\frac{\kappa}{r}$.

(Notice that these two ``polarizations" correspond to the two inequivalent
two by two irreducible representations of the gamma
matrices\cite{Flekkoy}).

The general solution to Eq.~(\ref{hache}) can be written as a
combination of
\be
\Psi_E^{(I)} = \left( \begin{array}{c} \psi_E^+ \\ 0
\end{array}
\right), \qquad \Psi_E^{(II)} = \left(
\begin{array}{c} 0 \\ \psi_E^- \end{array} \right)\,,
\ee with
\be
H_{\pm} \psi_E^{\pm} = E \psi_E^{\pm} \,.\label{plusminus} \ee

After separating variables, and for noninteger $\kappa =\ell+a$ (where
$\ell$ is the integer part of the reduced flux, and $a$ its fractionary
part), the eigenfunctions in Eq.~(\ref{hache}) turn out to be

\[
\Psi_E\left(r,\theta\right) = \left(
\begin{array}{l}
\sum_{n=-\infty}^\infty f_n^{+}(r)e^{in\theta}\\
\sum_{n=-\infty}^\infty g_n^{+}(r)e^{i\left(n+1\right)\theta}\\
\sum_{n=-\infty}^\infty f_n^{-}(r)e^{in\theta}\\
\sum_{n=-\infty}^\infty g_n^{-}(r)e^{i\left(n+1\right)\theta}
\end{array}
\right)\]
\begin{equation}
=\left(
\begin{array}{l}
\sum_{n=-\infty}^\infty \left(A_n^+\,J_{n-\kappa}\left(k r\right)+
B_n^+\,J_{\kappa-n}\left(k r\right)\right) e^{i n\theta} \\
\sum_{n=-\infty}^\infty
-i\frac{k}{E-m}\left(A_n^+\,J_{n+1-\kappa}\left(k r\right)-
B_n^+\,J_{\kappa-n-1}\left(k r\right)\right) e^{i
\left(n+1\right)\theta}\\ \sum_{n=-\infty}^\infty
\left(A_n^-\,J_{n-\kappa}\left(k r\right)+
B_n^-\,J_{\kappa-n}\left(k r\right)\right) e^{i n\theta} \\
\sum_{n=-\infty}^\infty
-i\frac{k}{E+m}\left(A_n^-\,J_{n+1-\kappa}\left(k r\right)-
B_n^-\,J_{\kappa-n-1}\left(k r\right)\right) e^{i
\left(n+1\right)\theta}
\end{array}
\right)\,, \label{eq-18}
\end{equation}

where $k= + \sqrt{(E^2-m^2)}$.

(Of course, for integer $\kappa$, a linear combination of Bessel and
Neumann functions must be taken).

\section{Behaviour at the origin}
\label{sec-origen}

As is well known \cite{AlRuWil1989,GerbertJackiw,Gerbert}, the radial
Dirac Hamiltonian in the background of an Aharonov-Bohm gauge field
requires a self-adjoint extension for the critical subspace $n=\ell$. In
fact, imposing regularity of all components of the Dirac field at the
origin is too strong a requirement, except for integer flux. Rather, one
has to apply the theory of Von Neumann deficiency indices \cite{Reed},
which leads to a one-parameter family of allowed boundary conditions
\cite{Gerbert}, characterized by
\begin{equation}
i\,\lim_{r\rightarrow 0}\left(m r\right)^{1-a} g_{\ell}^{\pm}(r)
\sin\left(\frac{\pi}{4}+\frac{\Theta^{\pm}}{2}\right)=
\lim_{r\rightarrow 0}\left(m r\right)^{a} f_{\ell}^{\pm}(r)
\cos\left(\frac{\pi}{4}+\frac{\Theta^{\pm}}{2}\right)\,.
\end{equation}
Here, $\Theta^{\pm}$ parameterize the admissible self-adjoint extensions
of $H_{\pm}$ respectively. Which of these extensions to choose depends on
the physical situation under study.

Throughout this paper we will, for non-integer $\kappa$, consider two
different behaviours at the origin. The first one, from now on called
behaviour I, is characterized by
\begin{equation}
\Theta^{\pm} =\left\{\begin{array}{rl}
-\frac{\pi}{2} & \quad\text{for}\, a\geq\frac{1}{2} \\
\frac{\pi}{2} & \quad \text{for}\, a<\frac{1}{2}
\end{array}\right.\,.
\end{equation}

As shown in Refs.~\cite{nos1,nos2}, this is the extension arising
when boundary conditions of the Atiyah-Patodi-Singer (APS) type
\cite{aps,aps2,aps3,spectral,Ma,Ninomiya} are imposed at a finite radius,
then taken to zero.

The second self-adjoint extension we will consider, from now on called
behaviour II, corresponds to
\begin{equation} \Theta^{\pm} =\left\{\begin{array}{rl} -\frac{\pi}{2} &
\quad \text{for}\, \kappa>0 \\ \frac{\pi}{2} & \quad \text{for}\, \kappa<0
\ \end{array}\right.\,. \label{hagen} \end{equation}

As shown in Ref.~\cite{Hagen}, this extension arises when a finite
radius flux tube is considered, thus asking for continuity of the
components of the spinor, and then shrinking the radius to zero.

Outside the critical subspace, the eigenfunctions in Eq.~(\ref{eq-18}) are
determined by the requirement of square integrability at
the origin, and are thus identical for the behaviours I and II. They are
given by
\begin{equation}
\Psi_E^{n\leq \ell-1}\left(r,\theta\right)= \left(\begin{array}{l}
B_n^+ J_{\ell+a-n}\left(k r\right) e^{i n \theta} \\
i\frac{k}{E-m}B_n^+ J_{\ell+a-n-1}\left(k r\right)
e^{i\left(n+1\right) \theta}\\ B_n^- J_{\ell+a-n}\left(k r\right)
e^{i n \theta} \\ i\frac{k}{E+m}B_n^- J_{\ell+a-n-1}\left(k
r\right) e^{i\left(n+1\right) \theta}
\end{array}\right)
\end{equation}
and
\begin{equation}
\Psi_E^{n\geq\ell+1}\left(r,\theta\right)= \left(\begin{array}{l}
A_n^+J_{n-\ell-a}\left(k r\right) e^{i n \theta}
\\ -i\frac{k}{E-m}A_n^+J_{n+1-\ell-a}\left(k r\right)
e^{i\left(n+1\right) \theta}\\ A_n^-J_{n-\ell-a}\left(k r\right)
e^{i n \theta} \\ -i\frac{k}{E+m}A_n^-J_{n+1-\ell-a}\left(k
r\right) e^{i\left(n+1\right) \theta} \end{array}\right)\,.
\label{eq-22} \end{equation}

In the critical subspace ($n=\ell$), the eigenfunction for behaviour I is
given by
\begin{equation}
\Psi_E^{\ell}\left(r,\theta\right)= \left(\begin{array}{l}
B_\ell^+ J_{a}\left(k r\right) e^{i \ell \theta}
\\ i\frac{k}{E-m}B_\ell^+ J_{a-1}\left(k r\right)
e^{i\left(\ell+1\right) \theta}\\ B_\ell^- J_{a}\left(k r\right)
e^{i \ell \theta} \\ i\frac{k}{E+m}B_\ell^- J_{a-1}\left(k
r\right) e^{i\left(\ell+1\right) \theta}
\end{array}\right)\qquad \text{for} \,a\geq\frac12, \label{cr1}
\end{equation}
and
\begin{equation}
\Psi_E^{\ell}\left(r,\theta\right)= \left(\begin{array}{l}
A_\ell^+ J_{-a}\left(k r\right) e^{i \ell \theta}
\\ -i\frac{k}{E-m}A_\ell^+ J_{1-a}\left(k r\right)
e^{i\left(\ell+1\right) \theta}\\ A_\ell^- J_{-a}\left(k r\right)
e^{i \ell \theta} \\ -i\frac{k}{E+m}A_\ell^- J_{1-a}\left(k
r\right) e^{i\left(\ell+1\right) \theta}
\end{array}\right)\qquad \text{for} \,a<\frac12\,. \label{cr2}
\end{equation}

It is easy to see that this extension satisfies the condition of minimal
irregularity (the radial functions diverge as $r\rightarrow 0$ at most as
$r^{-p}$, with $p\leq \frac{1}{2}$). Moreover, it is compatible with
periodicity in $\kappa$, a natural requirement when the origin is an
excluded point.

When behaviour II is imposed at the origin, the eigenfunctions in the
critical subspace are given by Eq.~(\ref{cr1}) for $\kappa>0$, and by
Eq.~(\ref{cr2}) for $\kappa<0$.

It is worth pointing out that, for integer $\kappa=\ell$, both APS
boundary conditions and the finite radius flux tube lead, when taking the
singular limit, to the requirement of regularity of all components at the
origin. In this case
\begin{equation}
\Psi_E\left(r,\theta\right) =  \sum_{n=-\infty}^\infty
\left(\begin{array}{l} A_n^+J_{n-\kappa}\left(k r\right) e^{i
n\theta} \\ -i\frac{k}{E-m}A_n^+ J_{n+1-\kappa}\left(k
r\right)e^{i \left(n+1\right)\theta}\\ A_n^-J_{n-\kappa}\left(k
r\right) e^{i n\theta} \\ -i\frac{k}{E+m}A_n^-
J_{n+1-\kappa}\left(k r\right)e^{i \left(n+1\right)\theta}
\end{array} \right)\,. \label{eq-25}
\end{equation}

\section{The theory in a bounded region. Energy spectrum and Casimir
energy} \label{sec-bounded}

From now on, we will confine the Dirac fields inside a bounded region, by
introducing a boundary at $r=R$, and imposing there MIT bag boundary
conditions.

The Casimir energy is formally given by
\begin{equation}
E_C = -\frac{1}{2} \left( \sum_{E>m} E_{\rho} - \sum_{E<-m}
E_{\rho} \right)\,, \label{eq-ec}
\end{equation}
where $\rho$ represents all indices appearing in the eigenvalue equation
that arises after local MIT conditions are imposed. In doing so, one must
consider a boundary operator $B$ which, with the representation of the
Dirac matrices given in Eq.~(\ref{dirac}), is also block-diagonal, and can
be
written as
\be
B = 1 - i \not \! n = 1 + i (\gamma^1 n^1 + \gamma^2 n^2) = \left(
\begin{array}{cc} B_+ & 0_{2\times2} \\ 0_{2\times2} & B_-
\end{array} \right)\,,
\ee
where $n$ is the exterior normal, and
\be
B_{\pm} = \left( \begin{array}{cc} 1 & \pm i e^{-i \theta} \\ \mp i
e^{i \theta} & 1 \end{array} \right)\,. \ee

Consider, in the first place, behaviour I (II) for $a\geq 1/2$
($\kappa>0$). Then, the eigenvalue equations for the upper (+)
polarization are
\begin{equation}
J_{n+a}(kR) = \frac{k}{E-m} J_{n-1+a}(kR) \qquad \text{for} \,\,
n=1,...,\infty, \label{r3}\end{equation}
\begin{equation}
J_{n-a}(kR) = \frac{-k}{E-m} J_{n+1-a}(kR) \qquad \text{for}\,
\,n=1,...,\infty \,,\label{r4}
\end{equation}
coming from non-critical subspaces, and
\begin{equation}
J_{a}(k R) = \frac{k}{E-m} J_{a-1}(k R)\,,
\label{r5}\end{equation}
which comes from the critical one.

The eigenvalue equations corresponding to the lower ($-$) polarization are
\begin{equation}
J_{n+a}(k R) = \frac{-k}{E+m} J_{n-1+a}(k R) \qquad \text{for}\,
\, n=1,...,\infty, \label{r1}
\end{equation}
\begin{equation}
J_{n-a}(k R) = \frac{k}{E+m} J_{n+1-a}(k R) \qquad \text{for}\,
\,n=1,...,\infty\,, \label{r2}
\end{equation}
from non-critical subspaces, and
\begin{equation}
J_{a}(k R) = \frac{-k}{E+m} J_{a-1}(k R)\,,
\label{r6}\end{equation}
which comes from the critical one.

For $a<1/2$ ($\kappa<0$), the contributions from non-critical subspaces
are the same, while those due to $n=\ell$ are
\begin{equation}
J_{-a}(k R) = \frac{-k}{E-m} J_{1-a}(k R)\,,
\label{r7}\end{equation} for the upper polarization, and
\begin{equation}
J_{-a}(k R) = \frac{k}{E+m} J_{1-a}(k R)\,,
\label{r8}\end{equation}
for the lower one.

It is easy to verify that positive energies coming from one polarization
correspond to negative energies coming from the other. Thus, both
polarizations give identical contributions to the Casimir energy in
Eq.~(\ref{eq-ec}).

Then, the formal expression for the Casimir energy (\ref{eq-ec}) is
\begin{equation}
E_C =-\frac{1}{2}2\sum_{k}{\left(k^2+m^2\right)}^{1/2}\,,
\label{eq-ec2}
\end{equation}
where the $k$'s are solutions of
\begin{mathletters}
\label{eq-raices}
\begin{eqnarray}
J^2_{n+a}(k R) - J^2_{n-1+a}(k R) - \frac{2m}{k} J_{n+a}(k
R)J_{n-1+a}(k R)& = & 0  \qquad \text{for}\, \, n=0,...,\infty,
\\ J^2_{n-a}(k R) - J^2_{n+1-a}(k R) + \frac{2m}{k} J_{n-a}(k
R)J_{n+1-a}(k R) & = & 0 \qquad \text{for}\, \,n=1,...,\infty,
\end{eqnarray}
\end{mathletters}
when $a\geq 1/2$ ($\kappa>0$) while, for $a<1/2$ ($\kappa<0$), the first
equation in (\ref{eq-raices}) holds for $n=1,...,\infty$ and the second
one applies for $n=0,...,\infty$.

\bigskip

Of course, a regularization method must be introduced in order to give
sense to the divergent sum in Eq.~(\ref{eq-ec2}). In the framework of
the $\zeta$-regularization \cite{zeta,Dowker:1976tf} (for several
applications see \cite{eliz-book1,eliz-book2}),
\begin{equation}
E_C =-\frac{1}{2} M \lim_{s\rightarrow -\frac12} M^{2 s} \zeta (s)
= - \frac12 M \lim_{s\rightarrow -\frac12} \, 2 \sum_{k}
\left(\frac{k^2+m^2}{M^2}\right)^{-s}\,, \label{eq-33}
\end{equation}
where the parameter $M$ was introduced for dimensional reasons.

Here, it is useful to define the so-called partial zeta function
\begin{equation}
\zeta_\mu (s) =  2 \sum_{l=1}^\infty \left( k_{\mu,l}^2 + m^2
\right)^{-s}\,,
\end{equation}
where $k_{\mu,l}$ are the roots of
\begin{equation}
 J^2_{\mu}(k R) - J^2_{\mu-1}(k R) - \frac{2m}{k}
  J_{\mu}(k R)J_{\mu-1}(k R)=0.
\end{equation}

So, after introducing $\nu= n+\frac12$ and $\alpha=a-\frac12$, the
Casimir energy for the behaviour I at the origin can be written,
for any $a$, as
\be
E_C^{I} = -\frac12 M \lim_{s\rightarrow -\frac12} M^{2s} \left\{
\sum_{\nu=\frac12,\frac32,\dots} \big[ \zeta_{\nu + \alpha} (s) +
\zeta_{\nu - \alpha} (s) \big] - \zeta_{\frac12 - |\alpha|}(s)
\right\}\,, \label{regu1} \ee
while for the behaviour II at the origin, it is given by
\be
E_C^{II} = -\frac12 M \lim_{s\rightarrow -\frac12} M^{2s} \left\{
\sum_{\nu=\frac12,\frac32,\dots} \big[ \zeta_{\nu + \alpha} (s) +
\zeta_{\nu - \alpha} (s) \big] - \zeta_{\frac12 -
\text{sgn}(\kappa)\,\alpha}(s) \right\}\,. \label{regu2} \ee

From Eq.~(\ref{regu1}) it is clear that, as mentioned before, the
Casimir energy for a behaviour of type I at the origin is independent of
the integer part $\ell$ of the reduced flux. Moreover, it is invariant
under $\alpha\rightarrow -\alpha$ ($a\rightarrow 1-a$). Thus, it is enough
to study it for $0<\alpha<\frac12$, where the absolute value in the last
term can be ignored, and to use $E_C^I (\alpha) = E_C^I (-\alpha)$.

Similarly, from Eq.~(\ref{regu2}), the Casimir energy for behaviour
II is seen to be invariant under $\kappa\rightarrow -\kappa$. Thus we only
study the case $\kappa>0$, where the last term is again $ \zeta_{\frac12 -
\alpha}(s)$ and $-\frac{1}{2} <\alpha<\frac12$ is considered.

\section{Evaluation of the Casimir energy}
\label{sec-eval}

The Casimir energies in Eqs.~(\ref{regu1}) and (\ref{regu2}) contain
two contributions: the term inside the square brackets, which is summed
over $\nu$, and the last term, which is a partial zeta function.

In both cases, it is useful to introduce, as in references
\cite{Leseduarte1996,Romeo1993,Romeo1994}, an integral representation for
the partial zeta function:
\be
\zeta_\mu (s) = 2 \frac{\sin \pi s}{\pi} R^{2s} \int_{z}^\infty dx \,
\left[ x^2 - z^2 \right]^{-s } \frac{d}{dx}
\ln \left[ x^{-2 \left( \mu - 1 \right)} F_\mu \left(x \right)
\right]\,, \label{eq37} \ee
where
\be
F_\mu (x) = I_\mu^2 (x) + I_{\mu-1}^2 (x) + \frac{2z}{x} I_\mu
(x) I_{\mu-1} (x)\,, \label{implicit} \ee
which have to be summed according to Eqs.~(\ref{regu1}) and (\ref{regu2}).
Here, the dimensionless variable $z=mR$ has been introduced.

In order to identify the divergences and evaluate the finite parts of the
terms in Eqs. (\ref{regu1}) and (\ref{regu2}) an analytical continuation
of the zeta function $\zeta(s)$ to $s=-\frac12$ has to be constructed. A
method to do this has been developed in \cite{Bordag:1996gm} and for
details of the procedure we refer to this reference. For the part of the
zeta functions involving the angular momentum sum the method consists of
adding and subtracting several orders of the uniform Debye expansion of
Eq.~(\ref{implicit}) such as to make the sum as well as the integral
in Eqs.~(\ref{regu1}), (\ref{regu2}) and (\ref{eq37}) well defined in an
increasing strip of the complex $s$-plane. For the partial zeta function,
instead, subtracting and adding the asymptotic terms for large arguments
of the Bessel functions will be sufficient.

Let us first study the terms summed over $\nu$. By making use of
the recurrence relations for Bessel functions, it is immediate to
obtain
\[
T(\mu,x,z) = \frac{d}{dx} \ln \left[ x^{-2\left( \mu - 1 \right)}
F_\mu (x) \right] =
\]
\be
\frac{2}\mu \left(\frac\mu{x}\right) \frac{1 + z - \left(
\frac{\mu}{x} \right)^2 z + 2 d_\mu (x) + \frac1{\mu^2}
\left(\frac{\mu}{x}\right)^2 z \, d_\mu^2(x)}{1 + \left(
\frac{\mu}{x} \right)^2 + \frac2{\mu} \left( \frac{\mu}{x}
\right)^2 z + \frac2\mu \left( \frac{\mu}{x} \right)^2 \, d_\mu
(x) + \frac2{\mu^2} \left( \frac{\mu}{x} \right)^2 z \, d_\mu (x)
+ \frac1{\mu^2} \left( \frac{\mu}{x}\right)^2 \, d_\mu^2 (x)}\,,
\ee where $ d_\mu (x) = x \frac{d}{dx} \ln I_\mu (x)$.
This expression can be developed in powers of $\frac1\mu$, through the
Debye expansion of Bessel functions, after taking $\left( \frac{\mu}{x}
\right)= \frac{\tau}{\sqrt{1-\tau^2}}$, with $\tau$ the variable of the
recursive polynomials $u_k (\tau)$ \cite{abram}.

If $D^{(N)} (\mu,x,\tau,z)$ is such an expansion up to the order
$\frac1{\mu^N}$, the partial zeta function can be written as
\be
\zeta_\mu (s) = \zeta_\mu^{a} (s) + \zeta_\mu^{d} (s)\,, \ee where
\be
\zeta_\mu^{a} (s) = 2 \frac{\sin \pi s}{\pi} R^{2s}
\int_{z}^{\infty} dx \, \left[x^2 - z^2 \right]^{-s} \left[
T(\mu,x,z) - D^{(N)} (\mu,x,\tau,z) \right]\label{asin} \ee
is the analytic part of the partial zeta function for $s=-\frac12$, while
\be
\zeta_\mu^{d} (s) = 2 \frac{\sin \pi s}{\pi}
R^{2s}\int_{z}^{\infty} dx \, \left[x^2 - z^2 \right]^{-s} D^{(N)}
(\mu,x,\tau,z)\, \ee
is the asymptotic contribution.

In order to make the integral in Eq.~(\ref{asin}), and the subsequent
sum over $\nu$, absolutely convergent at $s=-\frac12$, it is necessary to
take $N\geq2$ \cite{Bordag:1996gm}. We will choose $N=4$ to improve the
convergence of the sum of the analytic term (\ref{asin}), thus decreasing
the computational time needed to get accurate numerical results.

Now, the term in square brackets in Eqs.~(\ref{regu1}) and
(\ref{regu2}) involves the combination $\zeta_{\nu + \alpha} (s) +
\zeta_{\nu - \alpha} (s)$. In order to use previous results of
\cite{Elizalde:1997we} we further expand in powers of $\frac1{\nu}$. We
introduce the corresponding combination of asymptotic expansions,
\begin{eqnarray}
\Delta^{(N)} (\nu,x,t,z) & = & D^{(N)} \left( \nu \left( 1 +
\frac{\alpha}{\nu} \right),x,t \left( 1 + \frac{\alpha}{\nu}
\right) \left[ 1 + \frac{2 \alpha t^2}{\nu} \left( 1 +
\frac{\alpha}{2 \nu} \right) \right]^{-\frac12} , z \right) +
\nonumber \\ & & D^{(N)} \left( \nu \left( 1 - \frac{\alpha}{\nu}
\right),x,t \left( 1 - \frac{\alpha}{\nu} \right) \left[ 1 -
\frac{2 \alpha t^2}{\nu} \left( 1 - \frac{\alpha}{2 \nu} \right)
\right]^{-\frac12} , z \right)
\end{eqnarray}
expanded up to the order $\frac1{\nu^N}$. In the above expression $t =
\frac{\nu}{\sqrt{\nu^2 + x^2}}$

The asymptotic expansion can be written as
\be
\Delta^{(N)} (\nu,x,t,z) = \Delta_{-1} + \Delta_{0} +
\sum_{i=1}^{N} \Delta_i \,,\ee
where
\be
\Delta_{-1} = \frac{4 x}{\nu} \frac{t}{1+t}, \qquad \Delta_0 =
\frac{2x}{\nu^2} \frac{t^2}{1+t}, \qquad \Delta_i = \frac1{\nu^{i}}
\frac{d}{dx} \sum_{j=0}^{2i} b_{(i,j)} \, t^{i+j}\,. \label{eq-45}\ee
and the coefficients $b_{(i,j)}$ are listed in Appendix \ref{coeffs}.

Then, the term inside the square brackets in Eqs.~(\ref{regu1}) and
(\ref{regu2}) can be written as
\be
\lim_{s\rightarrow-\frac12} M^{2s} \sum_{n=0}^{\infty} \left[
\zeta_{\nu + \alpha} (s) + \zeta_{\nu-\alpha} (s) \right] =
Z(-\frac12) + \lim_{s\rightarrow-\frac12} M^{2s} \left( A_{-1}(s)
+ A_0 (s) + \sum_{i=1}^{N-1} A_i (s) \right)\,, \ee
where
\be
Z\left(-\frac12\right) = - \frac{2}{\pi M R} \sum_{n=0}^{\infty}
\int_{z}^\infty dx \, \left[ x^2 - (z)^2 \right]^{\frac12} \left\{
T \left( \nu + \alpha, x, z \right) + T\left( \nu- \alpha, x, z
\right) - \Delta^{(N)} \left(\nu,x,t,z \right) \right\} ,
\ee
\be
A_{-1}(s) = 2 \frac{\sin \pi s}{\pi} R^{2s} 4 \sum_{n=0}^{\infty}
\nu \int_{\frac{z}{\nu}}^\infty dx \, \left[ (\nu x)^2 - z^2
\right]^{- s } \frac{\sqrt{1+x^2} -1}{x}, \label{ec-am1}
\ee
\be
A_{0}(s) = 2 \frac{\sin \pi s}{\pi} R^{2s} 2 \sum_{n=0}^{\infty}
\int_{\frac{z}{\nu}}^\infty dx \, \left[ ( \nu x )^2 - z^2
\right]^{- s } \frac{1}{\sqrt{1+x^2}} \frac{\sqrt{1+x^2} -1}{x},
\label{ec-a0}
\ee
\be
A_{i}(s) = 2 \frac{\sin \pi  s }{\pi} R^{2s} \sum_{j=0}^{2i}
b_{(i,j)} \sum_{n=0}^{\infty} \nu^{-i}
\int_{\frac{z}{\nu}}^{\infty} dx \, \left[ (\nu x )^2 - z^2
\right]^{- s } \frac{d}{dx} \left( \frac{1}{\sqrt{1+x^2}}
\right)^{i+j}\,. \label{ec-ai} \ee
Eqs.~(\ref{ec-am1}), (\ref{ec-a0}) and (\ref{ec-ai}) can be expressed
in a systematic way by introducing the functions \cite{Elizalde:1997we}
\be
f(s;a,b;x) = \sum_{n=0}^{\infty} \nu^a \left[ 1 + \left(
\frac{\nu}{x} \right)^2 \right]^{-s-b}\,, \label{ec-efes}\ee
studied further in Appendix \ref{funciones-f}, which allow one to write
the asymptotic parts as
\be
A_{-1}(s)= \frac{2R^{2s}}{\sqrt{\pi}} \frac{\Gamma \left( s -
\frac12 \right) }{\Gamma(s)} z^{-2s+1} \int_0^1
\frac{dy}{\sqrt{y}} f\left(s;0,-\frac12; z\sqrt{y}\right),
\label{res-m1}
\ee
\be
A_0 (s) = \frac{2 R^{2s}}{\sqrt{\pi}} \frac{\Gamma \left( s +
\frac12 \right) }{\Gamma (s)} z^{-2s-1} \int_0^1
\frac{dy}{y^{\frac{3}{2}}} f\left(s;1,\frac12;z\sqrt{y}\right),
\label{res-0}
\ee
\be A_{i} (s) = \sum_{j=0}^{2i} b_{(i,j)} {\cal A}_{(i,j)} (s)
\,,\ee where \be {\cal A}_{(i,j)} (s) = - 2 R^{2s} z^{-(i+j)}
\frac{ \Gamma \left(s+ \frac{i+j}{2}\right) }{\Gamma
\left(\frac{i+j}{2} \right) \Gamma (s)} z^{-2s} f \left(
s;j,\frac{i+j}{2};z \right)\,. \label{res-i}
\ee

The complete expressions for these asymptotic parts around $s=-\frac12$
are derived in Appendix \ref{Aes}. Here, we list their residues, which
will be relevant to the discussion of the renormalization in the next
section:
\begin{mathletters}
\label{ec-Res}
\begin{eqnarray}
\text{Res}|_{s=-\frac12}\,A_{-1} & = & 0,  \\
\text{Res}|_{s=-\frac12}\,A_{0} & = & \frac1R \left[
\frac{z^2}{\pi} + \frac1{12 \pi } \right],
\\\text{Res}|_{s=-\frac12}\,A_{1} & = & 0, \\
\text{Res}|_{s=-\frac12}\,A_{2} & = & \frac1R \left[ \frac1{64} -
\frac1{12 \pi} - \frac{\alpha^2}{\pi} - \frac{z}{4} +
\frac{z}{\pi} - \frac{z^2}{2}\right].
\end{eqnarray}
\end{mathletters}

\bigskip

Next, we study the partial-zeta contribution
\be
e_c = -\frac12 M \lim_{s\rightarrow -\frac12} M^{2s} \left[ -
\zeta_{\frac12 - \alpha} (s) \right] \,
\ee
to the Casimir energy in Eqs.~(\ref{regu1}) and (\ref{regu2}), and
for reasons explained we omit the absolute value in the index.

In order to isolate the singularities, it is enough to consider the three
leading terms in the asymptotic expansion of Bessel functions for large
arguments, which will be denoted by $L(\frac12 - \alpha,x,z)$ ; thus, the
partial zeta function can be written as
\be
\zeta_{\frac12 - \alpha} (s) = \zeta_{\frac12 - \alpha}^{ a} (s) +
\zeta_{\frac12 - \alpha}^{ d} (s) \,,\ee where
\be
\zeta_{\frac12 - \alpha}^{a} (s) = 2 \frac{\sin \pi s}{\pi} R^{2s}
\int_{z}^{\infty} dx \, \left[x^2 - z^2 \right]^{-s} \left[
T\left({\frac12 - \alpha},x,z\right) - L\left({\frac12 -
\alpha},x,z\right) \right], \ee
\be
\zeta_{\frac12 - \alpha}^{d} (s) = 2 \frac{\sin \pi s}{\pi} R^{2s}
\int_{z}^{\infty} dx \, \left[x^2 - z^2 \right]^{-s} L\left({\frac12 -
\alpha},x,z\right)\,. \ee

Now, the subtracted terms can be written as
\be
L\left(\frac12 - \alpha,x,z\right) = \delta_{0} + \delta_{1} +
\delta_{2}\,,
\ee
where
\be
\delta_{0} = 2, \qquad \delta_{1} = \frac1{x} 2\alpha, \qquad
\delta_{2} = \frac1{x^2} (\alpha^2 - z) \,.\ee

Then
\be
\lim_{s\rightarrow-\frac12} M^{2s} \zeta_{\frac12 - \alpha} (s) =
z(-\frac12) + \lim_{s\rightarrow-\frac12} M^{2s} \left( a_{0} (s)
+ a_1 (s) + a_2 (s) \right)\,,
\ee
with
\be
z\left(-\frac12\right) = - \frac{2}{\pi M R} \int_{z}^\infty dx \,
\left[ x^2 - z^2 \right]^{\frac12} \left\{ T \left( \frac12 -
\alpha, x, z \right)  - L\left(\frac12 - \alpha,x,z \right)
\right\}\,,
\ee
which will be evaluated numerically, and
\be
a_0 (s) = \frac1R \left[ \frac{z^2}{\pi}
\frac{1}{\left(s+\frac12\right)} - \frac{z^2}{\pi} \left[ 1 + 2
\log \left(\frac{z}{2 R} \right) \right] + O\left(s+\frac12\right)
\right] ,
\ee
\be
a_1 (s) = \frac1R \left[ 2 \alpha z +
O\left(s+\frac12\right)\right] ,
\ee
\be
a_2 (s) = \frac1R (\alpha^2 - z) \left\{- \frac1{\pi}
\frac{1}{\left(s+\frac12\right)} + \frac1{\pi} \left[ 2 + 2 \log
\left( \frac{z}{2 R} \right) \right] + O\left(s+\frac12\right)
\right\}\,.
\ee

So, the residues at $s=-\frac12$ are given by
\begin{mathletters}
\label{ec-res}
\begin{eqnarray}
\text{Res}|_{s=-\frac12}\,a_0 & = & \frac1R
\left[\frac{z^2}{\pi}\right], \\ \text{Res}|_{s=-\frac12}\,a_1 & =
& 0, \\ \text{Res}|_{s=-\frac12}\,a_2 & = & \frac1R \left[ -
\frac{\alpha^2}{\pi } + \frac{z}{\pi}\right].
\end{eqnarray}
\end{mathletters}

\section{Discussion of the results}
\label{sec-discusion}

Clearly the Casimir energy is divergent and using
Eqs.~(\ref{ec-Res}) and Eqs.~(\ref{ec-res}) in Eqs.~(\ref{regu1})
and (\ref{regu2}), the total residue is given by
\be
\text{Res}|_{s=-\frac12} E_{C}= -\frac1{2 R} \left\{\frac1{64} -
\frac{z}{4} - \frac{z^2}{2}\right\}\,,
\ee
which is independent of the flux. Thus, the difference between Casimir
energies with arbitrary and integer flux is finite and contains the
relevant information about the effect of the flux.

\begin{figure} \epsffile{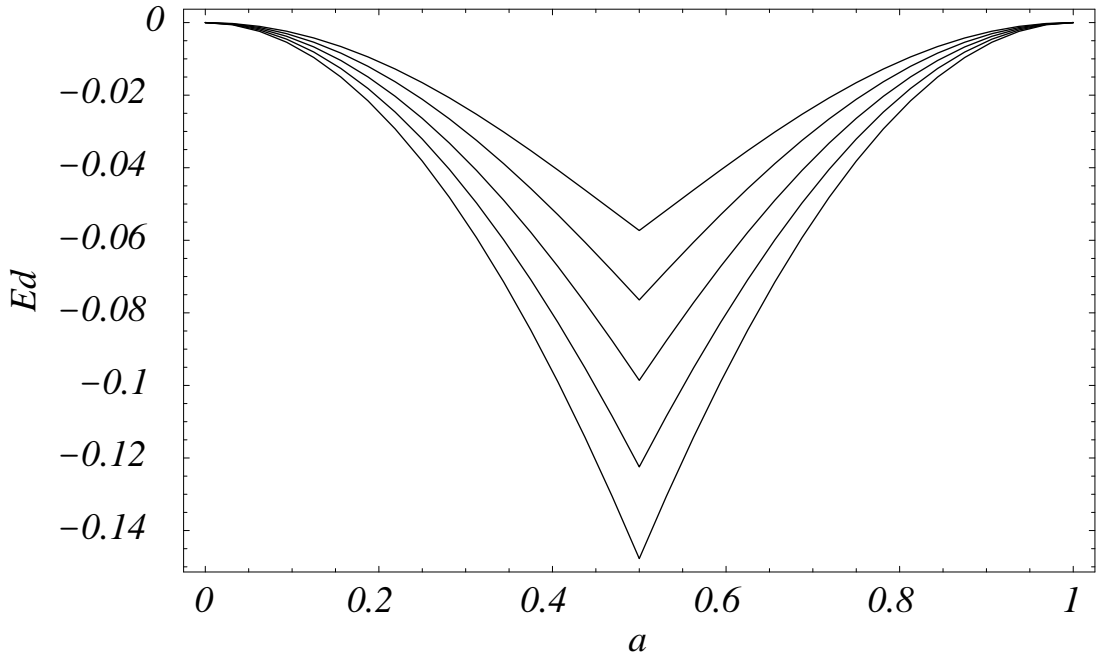} \caption{Difference $E_d$ of Casimir
energies. Behaviour I at the origin. Top to bottom: $z=
\frac{1}{128},\frac18, \frac14, \frac38, \frac{1}{2}$.} \label{fig-1}
\end{figure}

In Fig.~\ref{fig-1}, we plot the dimensionless difference $E_d= R
\left(E_C(a) - E_C(0)\right)$ for a behaviour of type I at the origin, as
a function of $a$ (the fractionary part of the reduced flux), for
different values of $z$. Since the finite part of the Casimir energy is
continuous in $a$, the difference goes to zero both at $a=0$ and $a=1$. It
shows a minimum at $a=\frac12$. It is interesting to note, that around
$a=0$, the vacuum energy decreases when the flux grows and that this
effect is enhanced with increasing mass.

\begin{figure} \epsffile{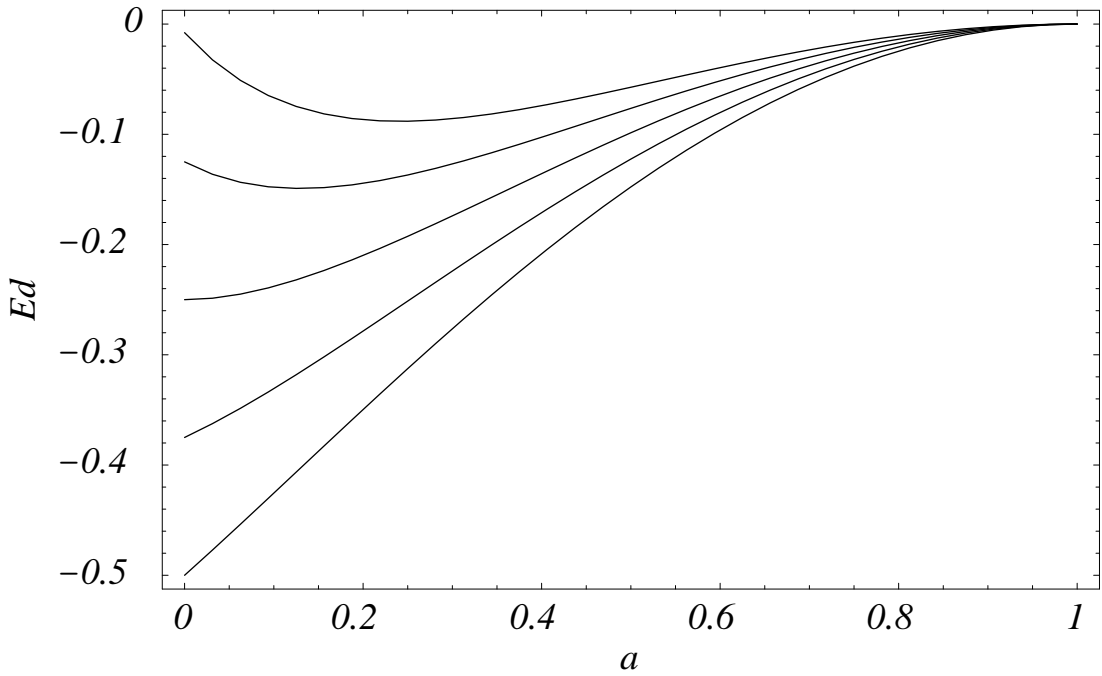} \caption{Difference $E_d$ of Casimir
energies. Behaviour II at the origin. Top to bottom:
$z=\frac{1}{128},\frac18, \frac14, \frac38, \frac{1}{2}$.} \label{fig-2}
\end{figure}

For a type II behaviour at the origin, the same difference is plotted in
Fig.~\ref{fig-2}. With decreasing mass our curves tend to the $m=0$ result
of Ref.~\cite{Leseduarte1996} and already $z=\frac1{128}$ shows quite a
good agreement with the corresponding figure in that reference (except for
a factor of 2, due to the fact that only one polarization was considered
by the authors of \cite{Leseduarte1996}).

Whereas for small values of the mass the energy exhibits a minimum
at $a \neq 0$, for larger values of $m$ this minimum is shifted
towards $a=0$. Furthermore, for $m \neq 0$ a nonzero value is seen
to arise for $a\rightarrow 0^{+}$. This is due to the
discontinuous behaviour of the finite part of the vacuum energy,
more precisely of the contribution of the critical subspace, at
integer values of the flux.

\begin{figure}
\epsffile{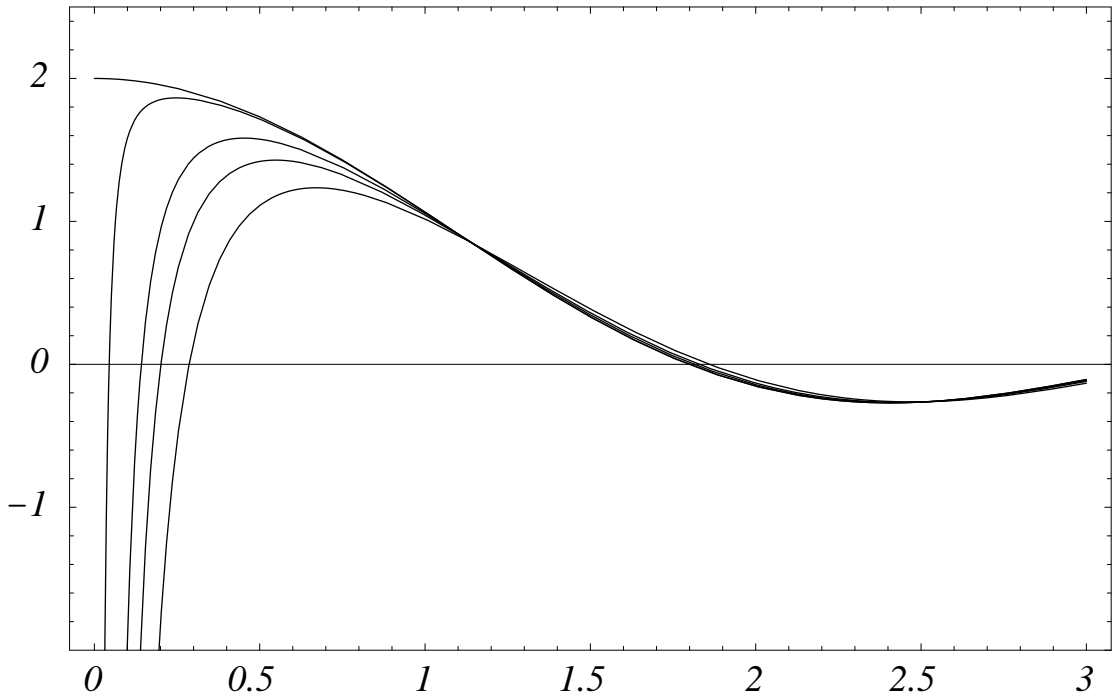} \caption{$J_a^2(kR)-J_{a-1}^2(kR)-\frac{2m}{k}
J_a(kR) J_{a-1}(kR)$. Top to bottom:$a=0, 0.001, 0.01, 0.02$}
\label{fig-3}
\end{figure}

The origin of this discontinuity can be traced back to the appearance, for
$a\rightarrow 0^{+}$, of a root of the combination of Bessel functions
involved in the partial zeta $\zeta_{a}$. Such a root is absent when
$a=0$. For $a\rightarrow 0^{+}$, this root goes to zero and, thus, gives
rise to a gap, which equals $m$. The quantity
$J_a^2(kR)-J_{a-1}^2(kR)-\frac{2m}{k} J_a(kR) J_{a-1}(kR)$ is shown in
Fig.~\ref{fig-3} as a function of $kR$, for various values of $a$ in order
to clarify this discontinuous behaviour.

In summary, we have seen that the presence of the mass as well as
the choice of the self-adjoint extension of the Hamiltonian have a
considerable influence on the dependence of the Casimir energy of
the flux. We have analyzed and discussed in detail the behavior of
the Casimir energy as a function of the parameters, namely flux
and mass. As a consequence of the mass, in the case of the
self-adjoint extension, (\ref{hagen}), a discontinuity of the
energy is found. It would be interesting to see how this effect
comes about, starting with the finite radius flux \cite{Hagen} and
shrinking the radius to zero. Does the effect persist or it is
only a result of the singular vortex? Furthermore, one should try
to understand better the physical meaning of the different
self-adjoint extensions by considering how the Casimir energy
depends on the parameter $\Theta$ of the one-parameter family of
self-adjoint extensions. Finally, more realistic
$(3+1)$-dimensional calculation should be envisaged.

\acknowledgements

We would like to thank Stuart Dowker and Horacio Falomir for
interesting discussions on the subject. KK has been supported by
the EPSRC under Grant No GR/M45726. CGB and EMS have been
supported by UNLP, under Grant No 11/X230, ANPCYT, under Grant
PICT'97 No 00039, and CONICET, under Grant PIP No 0459. MDF has
been supported by CONICET and UNLP.

\appendix

\section{Coefficients}
\label{coeffs}

In this Appendix we list the coefficients $b_{(i,j)}$ defined by
Eq.~(\ref{eq-45})
\begin{eqnarray} b_{(1,0)} & = & -{1\over 2} - 2\,{{\alpha }^2} + 2\,z
\nonumber \\ b_{(1,1)} & = & 0 \nonumber \\ b_{(1,2)} & = & {1\over 6}
\nonumber \\ b_{(2,0)} & = & -\left( z^2 \right) \nonumber \\ b_{(2,1)} &
= & -{1\over 4} - {{\alpha }^2} + z \nonumber \\ b_{(2,2)} & = & {1\over
4} - z \nonumber \\ b_{(2,3)} & = & {1\over 4} \nonumber \\ b_{(2,4)} & =
& -{1\over 4} \nonumber \\ b_{(3,0)} & = & {5\over {96}} + {{5\,{{\alpha
}^2}}\over {12}}+ {{{{\alpha }^4}}\over 6}-\left( {1\over 4} + {{\alpha
}^2} \right) z + {{2\,{z^3}}\over 3} \nonumber \\ b_{(3,1)} & = & -{1\over
4} + z - {z^2} \nonumber \\ b_{(3,2)} & = & {9\over {160}} - 2\,{{\alpha
}^2} - {{{{\alpha }^4}}\over 2} - \left( {1\over 2} - 3\,{{\alpha }^2}
\right) \,z + {z^2} \nonumber \\ b_{(3,3)} & = & 1 - 2\,z \nonumber \\
b_{(3,4)} & = & -{{23}\over {32}} + {{5\,{{\alpha }^2}}\over 4} +
{{7\,z}\over 4} \nonumber \\ b_{(3,5)} & = & -{3\over 4} \nonumber \\
b_{(3,6)} & = & {{179}\over {288}} \nonumber \\ b_{(4,0)} & = & {1\over
{16}} + {{{{\alpha }^2}}\over 4} - \left( {1\over 4} + {{\alpha }^2}
\right) \,z + \left( {1\over 4} + {{\alpha }^2} \right) \,{z^2} -
{{{z^4}}\over 2} \nonumber \\ b_{(4,1)} & = & -{{17}\over {64}} +
{{15\,{{\alpha }^2}}\over 8} + {{3\,{{\alpha }^4}}\over 4} - \left(
-{7\over 8} + {{9\,{{\alpha }^2}}\over 2} \right) \,z - {z^2} + {z^3}
\nonumber \\ b_{(4,2)} & = & -{1\over 4} - 4\,{{\alpha }^2} - \left(
-{1\over 2} - 10\,{{\alpha }^2} \right) \,z + \left( {1\over 4} -
4\,{{\alpha }^2} \right) \,{z^2} - {z^3} \nonumber \\ b_{(4,3)} & = &
{{165}\over {64}} - {{25\,{{\alpha }^2}}\over 4} - {{5\,{{\alpha
}^4}}\over 4} - \left( 6 - {{15\,{{\alpha }^2}}\over 2} \right) \,z +
{{5\,{z^2}}\over 2} \nonumber \\ b_{(4,4)} & = & -{{37}\over {32}} +
{{39\,{{\alpha }^2}}\over 4} - \left( -4 + 12\,{{\alpha }^2} \right) \,z -
2\,{z^2} \nonumber \\ x_{(4,5)} & = & -{{327}\over {64}} + {{35\,{{\alpha
}^2}}\over 8} + {{49\,z}\over 8} \nonumber \\ b_{(4,6)} & = & {{57}\over
{16}} - 6\,{{\alpha }^2} - {{21\,z}\over 4} \nonumber \\ b_{(4,7)} & = &
{{179}\over {64}} \nonumber \\ b_{(4,8)} & = & -{{71}\over {32}} \nonumber
\end{eqnarray}

\section{Functions \lowercase{$f(s;a,b;x)$}}
\label{funciones-f}

Here we are going to provide all analytical properties of the functions
$f(s;a,b;x)$ defined in Eq.~(\ref{ec-efes}).

As in Ref.~\cite{Elizalde:1997we}, we will make use of
\be
\sum_{n=0}^{\infty} h(\nu) = \int_0^{\infty} d\nu \, h(\nu) - i
\int_{0}^{\infty} d\nu \, \frac{h(i\nu +\epsilon) - h(-i\nu +
\epsilon )}{1+e^{2 \pi \nu}} \ee in the limit $\epsilon
\rightarrow 0$.

When applied to $h(\nu)=\nu^a \left( 1+ \left(\frac{\nu}{x}
\right)^2 \right)^{-t}$, the previous equation gives
\begin{eqnarray}
f(t;a,0;x)=\sum_{n=0}^{\infty} \nu^a \left( 1+ \left(
\frac{\nu}{x} \right)^2 \right)^{-t}
=
x^{a+1} & & \left\{ \frac12 \frac{\Gamma \left( \frac{a+1}{2}
\right) \Gamma \left( t - \frac{a+1}{2} \right)} {\Gamma (t)} +
\right.  \nonumber \\ & & 2 \sin \left( \frac{\pi a}{2} \right)
\int_{0}^{1} du \, \frac{u^a}{1+e^{2 \pi u x}} \left(1 - u^2
\right)^{-t} +
  \nonumber \\
& & \left. 2 \sin \left( \frac{\pi a}{2} - \pi t \right)
\int_{1}^{\infty} du \, \frac{u^a}{1+e^{2 \pi u x}} \left(u^2 - 1
\right)^{-t} \right\}\,. \label{suma}
\end{eqnarray}

Now, we are interested in $f(s;a,b;x)$, for arbitrary $b$. From the
definition, it is clear that $f(s;a,b;x) = f(s+b,a,0;x)$. However, as $b$
grows, the integrals in Eq.~(\ref{suma}) eventually diverge at $u=1$. In
order
to avoid such divergences, we will perform an adequate number of
integrations by parts, thus obtaining

\begin{eqnarray}
f(t;a,0;x) = & & \nonumber \\ x^{a+1} & &\left\{ \frac12 \frac{
\Gamma \left(\frac{a+1}{2} \right) \Gamma \left( t - \frac{a+1}{2}
\right)}{\Gamma \left( t \right)} - \right. \nonumber \\ & &
\left( \frac{e^{i \pi a}+1}{2} \right)
\frac{(-1)^{\frac{a}2}}{2^{\frac{a-2}{2}}} \sin (\pi t)
\frac{\Gamma \left( t - \frac{a}{2} \right) }{\Gamma (t)}
\int_1^\infty u \, du \, g\left( \frac{a}2 , a-1;u,x \right) (u^2
-1)^{-t + \frac{a}{2}} - \nonumber \\ & & \left. \left( \frac{e^{i
\pi a}-1}{2} \right) \frac{1}{2^{\frac{a-3}{2}}} \frac{\Gamma
\left( t - \frac{a-1}{2} \right) }{\Gamma (t)} \left[ \int_0^1 u
\, du \, g \left( \frac{a-1}2, a-1;u,x \right) (1-u^2)^{-t +
\frac{a-1}{2}} + \right. \right. \nonumber \\ & & \left. \left.
\phantom{\left( \frac{e^{i \pi a}-1}{2} \right) }
(-1)^{\frac{a-1}{2}} \cos (\pi t) \int_1^\infty u \, du \, g
\left( \frac{a-1}2, a-1;u,x \right) (u^2 -1)^{-t + \frac{a-1}{2}}
\right] \right\}\,,
\end{eqnarray}
where
\be
g(a,b;u,x) = \left( \frac1u \frac{d}{du} \right)^a
\frac{u^b}{1+e^{2 \pi u x}}\,.\label{ge} \ee

However, the number of integrations by parts is bounded by the
requirement that the integrated terms are well-behaved at $u=0$.
In what follows, we will thus keep the number of integrations
admissible, by making use of the following recurrence relationship
\be
f(s;a,b;x)= f(s;a,b-1;x) - \frac1{x^2} f(s;a+2,b;x)\,.
\label{recu} \ee

In this way, all the required functions can be reduced to four
different cases
\begin{itemize}
\item $f(s;2n,n;x)$, $f(s;2n,n+\frac12;x)$, $n=0,1,2,3,4,5,6,7$
\item $f(s;2n+1,n;x)$, $f(s;2n+1,n+\frac12;x)$, $n=0,1,2,3,4,5,6$\ .
\end{itemize}

Finally, after expanding in powers of $s+\frac12$, we get
\begin{eqnarray}
\lefteqn{f \left(2n,n;x \right) =}  \nonumber \\ x^{2n+1} & &
\left\{ -\frac12 \frac1{s+\frac12} \left(n-\frac12\right) -
\frac12 \left[ \left(n+\frac12\right)+
\left(n-\frac12\right)\left[ \psi(1) - \psi \left(
n+\frac12\right) \right] \right]- \right. \nonumber \\ & & \left.
\frac{\pi}{2^{n-2}} \frac{\left(n-\frac12\right)} {\Gamma
\left(\frac12\right)\Gamma \left( n + \frac12 \right)}
\int_1^\infty u \, du \, g(n,2n-1;u,x) \left( u^2 -1
\right)^{\frac12} + O(s+\frac12) \right\} \label{a6}
\\ \lefteqn{ f \left( 2n,n+\frac12;x \right) = } \nonumber \\
x^{2n+1} & &  \left\{ -n \frac{ \Gamma \left( n+ \frac12 \right)
\Gamma \left( \frac12 \right)}{\Gamma (n+1)} - n \frac1{2^{n-1}}
\frac{\pi}{\Gamma (n+1)} \int_1^\infty u\, du\, g(n,2n-1;u,x) +
O(s+\frac12) \right\}
\\
\lefteqn{f \left( 2n+1,n ;x \right) =}  \nonumber \\ x^{2(n+1)} &
& \left\{ \frac23 \frac{\Gamma (n+1)
\Gamma\left(\frac12\right)}{\Gamma \left(n+\frac12\right)}
\left(n-\frac12 \right) - \frac1{2^{n-2}}
\frac{\Gamma\left(\frac12\right)}{\Gamma \left(n+\frac12\right)}
\left(n-\frac12\right) \int_0^1 u\,du\, g(n,2n;u,x) \left( 1 - u^2
\right)^{\frac12} + O(s+\frac12) \right\} \\ \lefteqn{f \left(
2n+1,n+\frac12 ;x \right) =}  \nonumber \\ x^{2(n+1)} & & \left\{
-\frac1{2 s+\frac12} n \left[ 1 - \frac1{2^{n-2}} \frac1{\Gamma
(n+1)} \int_0^\infty u \, du\, g(n,2n;u,x) \right]- \frac12 \left[
1 + n + n \left[ \psi (1) - \psi (n+1)\right] \right] + \right.
\nonumber \\ & & \left.
 \frac1{2^{n-1}} \frac1{\Gamma (n+1)}
\left\{ \left[ 1 + n \left[ \psi (1) - \psi (n+1)\right] \right]
\int_0^\infty u\, du\, g(n,2n;u,x) \right. \right. \nonumber \\ &
& \left. \left. \phantom{ \frac1{2^{n-1}} \frac1{\Gamma (n+1)}} -n
\int_0^\infty u \, du \, g(n,2n;u,x) \ln \left| u^2 -1 \right|
\right\} + O(s+\frac12) \right\}\,, \label{a9}
\end{eqnarray}
where $\psi (x)$ is the Euler Psi function.

These expressions generate all the $f$-functions necessary  for
the evaluation of the required $A_i \left( s \right)$, for
$i=1,2,...$.

Notice that, for $j=0$ and $i=1$, the prefactor $\Gamma \left( s +
\frac{i+j}{2} \right)$ in Eq.~(\ref{res-i}) has a pole at
$s=-\frac12$. Thus, the order $s+\frac12$ in the expansion of
$f(s;0,\frac12;x)$ must be retained,

\be f\left( s;0,\frac12;x \right) = -\pi x
\left( s+\frac12 \right) \left[ 1 + 2  \int_1^\infty u\, du\, g(0,-1;u,x)
\right] +O\left((s+\frac12)^2\right) \,.\label{a10} \ee

\section{Evaluation of $A_{-1}$ and $A_0$}
\label{Aes}

Finally, in this Appendix we are going to describe some details of the
calculation of the $A_i (s)$, Eqs.~(\ref{res-m1})-(\ref{res-i}), at
$s=-\frac12$.

We will, in the first place, obtain an expression for $A_{-1}$ in
equation (\ref{res-m1})

\be
A_{-1}(s)= \frac{2R^{2s}}{\sqrt{\pi}} \frac{\Gamma \left( s -
\frac12 \right) }{\Gamma(s)} z^{-2s+1} \int_0^1
\frac{dy}{\sqrt{y}} f\left(s;0,-\frac12; z\sqrt{y}\right)
\label{A-1} \ee

By using equation (\ref{suma}) in the previous appendix, it can be
put in the form

\[
A_{-1}(s)=\frac{2R^{2s}}{\sqrt{\pi}} \frac{\Gamma \left( s -
\frac12 \right) }{\Gamma(s)} z^{-2s+1} \int_0^1
\frac{dy}{\sqrt{y}}\left\{ \frac12
\sqrt{y}\,z\frac{\Gamma(\frac12)\Gamma(s-1)}{\Gamma(s-\frac12)}\right.\]
\be
+ \left. 2\sqrt{y}\,z \sin\left(\pi(\frac12-s)\right)
\int_1^{\infty}du\,\left(u^2-1\right)^{\frac12-s}
\frac{1}{1+e^{2\pi u\sqrt{y}\,z}}\right\}. \ee

After interchanging the integrals, one gets
\[
A_{-1}(s)=\frac{R^{2s} z^{-2s+2}}{s-1} + \frac{4R^{2s}
z^{-2s+1}}{\sqrt{\pi}\Gamma(s)\Gamma(\frac32 -s)}\int_1^\infty
du\,\frac{\left(u^2-1\right)^{\frac12 -s}}{u}\ln\left(1+e^{-2\pi u
z }\right)
\]
\be
-\frac{2 R^{2s} z^{-2s}}{\sqrt{\pi}\Gamma(s)\Gamma(\frac32
-s)}\left\{ \frac{\pi^2 \Gamma(s)\Gamma(\frac32
-s)}{24\Gamma(\frac32 )}+ \int_1^\infty
du\,\frac{\left(u^2-1\right)^{\frac12 -s}}{u^2}\text{Li}_2\left(-e^{-2\pi
z u}\right)\right\}, \ee where
$\text{Li}_j(x)=\sum_{n=1}^{\infty}\frac{x^n}{n^{j}}$.

Finally, expanding around $s=-\frac12$, one has
\[
A_{-1} (s) = \frac{2 z^3}{R}\left[ - \frac13 -\frac1{12 z^2} +
\frac1{2 \pi^2 z^2} \int_1^\infty du \frac{u^2 -1}{u^2}\, \text{Li}_2
\left( - e^{- 2 \pi u z } \right) \right.
\]
\be \left. - \frac1{\pi z} \int_1^\infty du \frac{u^2 -1}{u} \log
\left( 1 + e^{- 2 \pi u z} \right) \right] + O(s+\frac12), \ee
which is a useful representation for numerical calculations.

Let us now go to the evaluation of $A_0$ in Eq.~(\ref{res-0}),
\be
A_0 (s) = \frac{2 R^{2s}}{\sqrt{\pi}} \frac{\Gamma \left( s +
\frac12 \right) }{\Gamma (s)} z^{-2s-1} \int_0^1
\frac{dy}{y^{\frac{3}{2}}} f\left(s;1,\frac12;z\sqrt{y}\right) \ee

As before, using Eq.~(\ref{suma}), and interchanging the
integration order, one gets
\[
A_0(s) = \frac{2R^{2s} z^{-2s-1}}{
\sqrt{\pi}}\frac{\Gamma(s+\frac12)}{\Gamma(s)}
\left\{\frac{z^2}{2\left(s-\frac12\right)}\int_0^1
\frac{dy}{y^{\frac12}} + 2z^2 \int_0^1
du\,u\left(1-u^2\right)^{-\left(s+\frac12\right)} \int_0^1 dy
\frac{1}{y^{\frac12}\left(1+e^{2\pi u z \sqrt{y}}\right)}\right.
\]
\be
\left.  +2\cos{\left(\pi \left(s+\frac12\right)\right)} z^2
\int_1^{\infty}du\,u\left(u^2
-1\right)^{-\left(s+\frac12\right)}\int_0^1 dy
\frac{1}{y^{\frac12}\left(1+e^{2\pi u z \sqrt{y}}\right)}\right\}
\ee

Now, after analytically extending, and developing around
$s=-\frac12$, one has
\[
A_0 (s) = \frac{z^2}{\pi R} \frac{1}{\left(s+\frac12\right)}
\left( 1 +\frac1{12 z^2} \right) - \frac{z^2}{\pi R} \left[ 1 +
\frac1{6z^2} + 2 \left( 1 + \frac1{12 z^2} \right) \log \left(
\frac{z}{2 R} \right) \right.
\]
\be
\left. + \frac2{\pi z} \int_0^\infty du \log |1-u^2| \log
\left(1+e^{- 2 \pi u z}\right) \right] + O(s+\frac12) \ee where
a simple pole at $s=-\frac12$ appears.

Clearly, in both cases the finite parts must be evaluated numerically. We
will not go into the detailed calculation of the $A_i$ for $i>0$, since it
is a direct consequence of the properties of $f(s;a,b;x)$ described in the
previous appendix.


\begin{thebibliography}{10}

\bibitem{Bohm}
Y. Aharonov and D. Bohm, Physical Review {\bf 115},  485  (1959).

\bibitem{AlRuWil1989}
M.~G. Alford, J. March-Russell, and F. Wilczek, Nuclear Physics {\bf B328},
  140  (1989).

\bibitem{Russel1988}
J. March-Russell and F. Wilczek, Physical Review Letters {\bf 61},  2066
  (1988).

\bibitem{AlWil1989}
M.~G. Alford and F. Wilczek, Physical Review Letters {\bf 62},  1071  (1989).

\bibitem{Hagen}
C.~R. Hagen, Phys. Rev. Lett. {\bf 64},  503  (1990).

\bibitem{Flekkoy}
E.~G. Flekk{\o}y and J.~M. Leinaas, International Journal of Modern Physics
  {\bf A6},  5327  (1991).

\bibitem{GerbertJackiw}
P. de~Sousa~Gerbert and R. Jackiw, Communications in Mathematical Physics {\bf
  124},  229  (1989).

\bibitem{Gerbert}
P. de~Sousa~Gerbert, Physical Review {\bf D40},  1346  (1989).

\bibitem{Niemi1984}
A.~J. Niemi and G.~W. Semenoff, Physical Review {\bf D30},  809  (1984).

\bibitem{Sitenko1996gd}
Y.~A. Sitenko, presented at 10th International Conference on Problems of
  Quantum Field Theory (Alushta 96) Alushta, Ukraine, 13-17 May 1996 -
  hep-th/9702148.

\bibitem{sitenko97}
Y.~A. Sitenko and D.~G. Rakityanskii, Phys. Atom. Nucl. {\bf 60},  1497
  (1997).

\bibitem{sitenko96}
Y.~A. Sitenko, Physics Letters {\bf B387},  334  (1996).

\bibitem{sitenkohep97}
Y.~A. Sitenko and D.~G. Rakityanskii, talk given at International Workshop on
  Mathematical Physics - 'Today, Priority Technologies - for Tomorrow', Kiew,
  Ucraine 12-17 May 1997 - hep-th/9710130.

\bibitem{Moroz}
A. Moroz, Physics Letters {\bf B358},  305  (1995).

\bibitem{fry}
M.~P. Fry, Physical Review {\bf D51},  810  (1995).

\bibitem{Bordag:1998}
M. Bordag and K. Kirsten, to appear in Physical Review D - hep-th/9812060.

\bibitem{bretvik}
I. Bretvik and T. Toverud, Classical Quantum Gravity {\bf 12},  1229  (1995).

\bibitem{parker}
L. Parker, Physical Review Letters {\bf 59},  1369  (1987).

\bibitem{Bordag:1991}
M. Bordag, Annals of Physics {\bf 206},  257  (1991).

\bibitem{scandurra}
M. Scandurra, Journal of Physics {\bf A32},  5679  (1999).

\bibitem{casimir}
H.~B.~G. Casimir, Proc. K. Ned. Acad. Wet. {\bf 51},  793  (1948).

\bibitem{Plunien}
G. Plunien, B. M{\"{u}}ller, and W. Greiner, Physics Reports {\bf 134},  87
  (1986).

\bibitem{nos1}
C.~G. Beneventano, M.~D. Francia, and E.~M. Santangelo, to appear in
  International Journal of Modern Physics A - hep-th/9809081.

\bibitem{nos2}
C.~G. Beneventano, M.~D. Francia, and E.~M. Santangelo,  in {\em Casimir energy
  - 50 years later}, Proceedings of the Fourth Workshop on Quantum Field Theory
  under the Influence of External Conditions, edited by M. Bordag (World
  Scientific, Leipzig, Germany, 1999), pp.\ 240--246.

\bibitem{Elizalde:1997we}
E. Elizalde, M. Bordag, and K. Kirsten, Journal of Physics {\bf 31},  1743
  (1998).

\bibitem{Leseduarte1996}
S. Leseduarte and A. Romeo, Commun. Math. Phys. {\bf 193},  317  (1998).

\bibitem{Bordag:1996ma}
M. Bordag, E. Elizalde, K. Kirsten, and S. Leseduarte, Physical Review {\bf
  D56},  4896  (1997).

\bibitem{manuel}
C. Manuel and R. Tarrach, Physics Letters {\bf B301},  72  (1993).

\bibitem{aps}
M.~F. Atiyah, V.~K. Patodi, and I.~M. Singer, Math. Proc. Camb. Phil. Soc. {\bf
  77},  43  (1975).

\bibitem{aps2}
M.~F. Atiyah, V.~K. Patodi, and I.~M. Singer, Math. Proc. Camb. Phil. Soc. {\bf
  78},  43  (1975).

\bibitem{aps3}
M.~F. Atiyah, V.~K. Patodi, and I.~M. Singer, Math. Proc. Camb. Phil. Soc. {\bf
  79},  71  (1976).

\bibitem{spectral}
H. Falomir, R.~E.~G. Sarav\'{\i}, and E.~M. Santangelo, Journal of Mathematical
  Physics {\bf 39},  532  (1998).

\bibitem{Ma}
Z.-Q. Ma, Journal of Physics {\bf A19},  L317  (1986).

\bibitem{Ninomiya}
M. Ninomiya and C.-I. Tan, Nuclear Physics {\bf B257},  199  (1985).

\bibitem{Poly}
A.~P. Polychronakos, Nuclear Physics {\bf B283},  268  (1987).

\bibitem{mit2}
P. Hasenfratz and J. Kuti, Physics Reports {\bf 40},  76  (1978).

\bibitem{zeta}
S.~W. Hawking, Communications in Mathematical Physics {\bf 55},  133  (1977).

\bibitem{Dowker:1976tf}
J.~S. Dowker and R. Critchley, Physical Review {\bf D13},  3224  (1976).

\bibitem{eliz-book1}
E. Elizalde {\it et~al.}, {\em Zeta regularization techniques with
  applications} (World Scientific, Singapore, 1994).

\bibitem{eliz-book2}
E. Elizalde, {\em Ten physical applications of spectral zeta functions}
  (Springer-Verlag, Berlin, 1995).

\bibitem{Romeo1993}
E. Elizalde, S. Leseduarte, and A. Romeo, Journal of Physics {\bf A26},  2409
  (1993).

\bibitem{Romeo1994}
S. Leseduarte and A. Romeo, Journal of Physics {\bf A27},  2483  (1994).

\bibitem{Bordag:1996gm}
M. Bordag, E. Elizalde, and K. Kirsten, J. Math. Phys. {\bf 37},  895  (1996).

\bibitem{Reed}
M. Reed and B. Simon, {\em {F}ourier {A}nalysis and {S}elf-{A}djointness}
  (Academic Press, New York, 1975).

\bibitem{abram}
M. Abramowitz and I. Stegun, {\em {H}andbook of {M}athematical {F}unctions}
  (Dover Publications, New York, 1970).

\end{thebibliography}

\end{document}